\documentclass[12pt,letterpaper]{article}

\usepackage[margin=1in]{geometry}
\usepackage{amsmath}
\usepackage{amsfonts}
\usepackage{amssymb}
\usepackage{braket}
\usepackage{graphicx}
\usepackage[colorlinks=true,linkcolor=blue,citecolor=blue,urlcolor=blue]{hyperref}
\usepackage{bm}
\usepackage{cite} 

\newcommand{\etal}{\textit{et al.}}
\newcommand{\ie}{\textit{i.e.}}
\newcommand{\eg}{\textit{e.g.}}
\newcommand{\Afield}{\mathcal A}
\newcommand{\bAfield}{\boldsymbol{\mathcal A}}

\begin{document}
	
	\title{Van der Waals interaction at short and long distances: \\
		from stationary perturbation theory to imaginary-time correlation 
		functions}
\author{L.~Saba$^{1}$ and C.~D.~Fosco$^{2}$\thanks{Corresponding author.}\\[2mm]
		\normalsize\textit{Centro At\'omico Bariloche and Instituto Balseiro,}\\
		\normalsize\textit{Comisi\'on Nacional de Energ\'{\i}a At\'omica,}\\
		\normalsize\textit{R8402AGP S.\ C.\ de Bariloche, Argentina}\\[2mm]
		\normalsize$^{1}$\texttt{lautaro.saba@ib.edu.ar} \quad
		$^{2}$\texttt{cesar.fosco@ib.edu.ar}}
	
	\date{}
	\maketitle
	\begin{abstract}
		The van der Waals interaction between neutral atoms is typically 
		studied using 
		stationary perturbation theory for the short-distance (London) limit, 
		while 
		long-distance (Casimir-Polder) results are usually derived via 
		semiclassical, 
		time-dependent approaches. Here, we demonstrate 
		that 
		reformulating stationary perturbation theory calculations in terms of 
		time-ordered correlation functions significantly simplifies the 
		mathematical 
		treatment. This reformulation is particularly advantageous for 
		higher-order 
		calculations required in the long-distance regime, where retardation 
		effects 
		become important. Our approach provides a unified framework connecting both 
		limiting cases, and is intended as a bridge between advanced quantum 
		mechanics and field-theoretic treatments of dispersion forces, suitable 
		for graduate-level courses or specialized readers.
	\end{abstract}
	
	
	\section{Introduction}
	\label{sec:intro}
	
	Van der Waals forces are responsible for phenomena as varied as the 
	adhesion of geckos to walls\cite{autumn2002evidence} and the stability of 
	biological
	membranes\cite{israelachvili2011intermolecular}. Their theoretical 
	description requires connecting two different regimes: short-distance 
	(London) and long-distance (Casimir-Polder), a task that remains 
	nontrivial from both conceptual and computational standpoints.
	
	The theoretical understanding of these forces developed over 
	several decades.  In 1873,
	van der Waals introduced his famous equation of state to account for 
	deviations
	from ideal gas behavior\cite{vanderwaalsdissertation}.
	However, the microscopic origin of these forces remained 
	mysterious until the advent of quantum mechanics. Keesom first 
	explained the interaction between molecules with permanent dipole 
	moments\cite{keesom1912}, but this could not account for forces between 
	nonpolar atoms.
	
	A breakthrough was made by London\cite{London1930}, who showed how 
	quantum fluctuations may induce forces between neutral atoms, even when 
	they 
	lack permanent dipole moments. London's theory predicts an interaction 
	energy 
	proportional to $R^{-6}$, where $R$ denotes the interatomic distance; this result is rederived below in Eq.~(\ref{eq:London}). His 
	treatment assumed an instantaneous (Coulomb) electromagnetic (EM) 
	interaction. However, this assumption is valid only 
	when $R \ll c/\Omega$, where $\Omega$ is a characteristic atomic frequency. 
	For larger separations, retardation effects become important. Casimir and 
	Polder addressed this issue in 1948\cite{Casimir1948}, showing that at 
	large distances the interaction transitions to a $R^{-7}$ law, 
	due to the finite speed of light; the corresponding asymptotic expression is recovered in Eq.~(\ref{eq:Casimir_Polder}). (The same year, Casimir also derived 
	the attraction between parallel conducting plates\cite{casimir1948attraction}, 
	which bears his name.) An alternative early derivation of retarded van der 
	Waals forces at all distances was later given by Boyer within stochastic 
	electrodynamics, using classical electromagnetic zero-point radiation\cite{Boyer1973}. 
	For a comprehensive modern treatment, 
	see Ref.~\cite{buhmann2012dispersion}. Note that the transition to the 
	$R^{-7}$ behavior by 
	no means implies that the system becomes time-dependent, the 
	Hamiltonian remains time-independent throughout.
	However, most derivations of the long-distance result rely upon a
	semiclassical approach where time dependence is introduced for the EM
	field.  
	
	It is one of our goals in this paper to show how a time-independent
	treatment of the system can be conveniently used to derive results for 
	both regimes. At the same time, we explain how an imaginary-time 
	correlation-function formulation greatly simplifies the inclusion of 
	retardation effects, which in stationary perturbation theory corresponds 
	to terms at fourth order in the perturbation. We also show how the same 
	framework helps study the London limit by including higher orders. The 
	presentation is aimed at readers who are already familiar with stationary 
	perturbation theory and have some exposure to field-theoretic methods. In 
	this sense, the paper may serve as complementary material in a graduate-level 
	quantum mechanics or quantum field theory course, rather than as a 
	self-contained undergraduate introduction.
	
	This paper is organized as follows: Section~\ref{sec:the_system} presents 
	the framework used to describe the system, introducing the
	atomic model and then the EM interaction.
	Section~\ref{sec:ham} presents the construction of the Lagrangian in the
	Coulomb gauge, used to evaluate corrections to the vacuum energy in
	the instantaneous approximation. We then 
	introduce a time-dependent treatment, showing how it simplifies the 
	evaluation
	of the vacuum energy in the dipole approximation to all orders in 
	perturbation theory. The stationary perturbation theory results are 
	summarily presented in an Appendix.
	
	Then, in Sec.~\ref{sec:ret}, we consider the inclusion of retardation 
	effects, showing how the time-dependent perturbation theory framework also
	simplifies the calculation. In Sect.~\ref{sec:conc} we present our 
	conclusions.
	
	
	\section{The system}
	\label{sec:the_system}
	
	The system we consider throughout most of this paper consists of
	two static, identical, neutral atoms, labeled $A$ and $B$, whose
	centers of mass occupy fixed positions $\mathbf{R}_A$ and
	$\mathbf{R}_B$, respectively. To an excellent approximation, which we
	adopt, each center coincides with the location of the respective nucleus.
	The positions of the least-bound electrons, measured from their own nuclei,
	are denoted by $\mathbf{r}_A$ and $\mathbf{r}_B$. Thus
	$\mathbf{R}_\alpha$ is a fixed nuclear position, whereas
	$\mathbf{r}_\alpha(t)$ is a dynamical electron coordinate. The interatomic
	separation is denoted by $R=|\mathbf{R}_A-\mathbf{R}_B|$. To avoid a clash
	between the atom label $A$ and the electromagnetic potential, we denote the
	latter by $\Afield_\mu=(\phi,\bAfield)$ throughout the paper.
	
	We take as our initial description the real-time action ${\mathcal S}$,
	assumed to have the form
	\begin{align}\label{eq:defs}
		{\mathcal S}( {\mathbf r}_A,{\mathbf r}_B , \Afield\,; {\mathbf R}_A ,
		{\mathbf R}_B )
		&=\;
		{\mathcal S}^a_0({\mathbf r}_A) + {\mathcal S}^a_0({\mathbf r}_B) +
		{\mathcal S}^a_I({\mathbf r}_A , \Afield\, ;{\mathbf R}_A ) \nonumber\\
		&+\, {\mathcal S}^a_I({\mathbf r}_B , \Afield\, ; {\mathbf R}_B)
		+ {\mathcal S}^{\rm EM}_0(\Afield)\;,
	\end{align}
	where ${\mathcal S}^a_0$ is the action for one electron in its binding
	potential, ${\mathcal S}^{\rm EM}_0(\Afield)$ is the action for the free EM
	field, and ${\mathcal S}^a_I$ contains the coupling of that field to an
	electron. Equation~(\ref{eq:defs}) is intended as a bookkeeping device: at
	this stage the model contains two atomic actions, one free-field action, and
	one atom-field interaction term for each atom. After gauge fixing or after
	introducing approximations, we shall explicitly state the corresponding
	effective Lagrangian before using it.
	
	Let us describe each of these terms in turn. The electrons have
	identical actions. Using $\alpha$ as a label running over the values
	$A$ or $B$, we take ${\mathcal S}^a_0$ to be 
	\begin{equation}\label{eq:defsah}
		{\mathcal S}^{a}_0({\mathbf r}_\alpha) \;=\;\frac{m}{2} 
		\int dt \,\Big( \dot{\mathbf{r}}_\alpha^2(t) - \Omega^2 \, {\mathbf
			r}_\alpha^2(t) \Big) \;,
	\end{equation}
	where $\Omega$ is an effective frequency describing the relevant
	dynamics of the electron in the atomic potential. This is the usual isotropic
	harmonic-oscillator model for an atom in dispersion-force calculations: it
	amounts to retaining the leading quadratic term in an effective binding
	potential and gives a dynamic polarizability with a single characteristic
	frequency. For standard treatments of this model in atom-field physics, see
	Refs.~\cite{cohen1989photons,buhmann2012dispersion}.
	
	Although simplified, this harmonic oscillator model yields the correct
	functional form for the polarizability and, as shown below, reproduces the
	$R^{-6}$ and $R^{-7}$ power laws. It is particularly appropriate
	for noble gas atoms, where the least-bound electron experiences an 
	approximately harmonic potential near equilibrium. Quantitatively 
	reasonable results are obtained when $\Omega$ is chosen to match 
	the atomic ionization energy.
	
	The EM field action, before gauge fixing, takes the well-known form
	\begin{equation}
		\mathcal{S}_{0}^{\text{EM}}(\Afield) = \frac{1}{2}\,
		\int dt \int d^3x \,\left({\mathbf E}^2(t,{\mathbf x}) -
		{\mathbf B}^2(t,{\mathbf x}) \right) = -\frac{1}{4} \,\int d^4x \,
		F_{\mu\nu}(x) F^{\mu\nu}(x)\;,
	\end{equation}
	with $d^4x \equiv dx^0\, dx^1\, dx^2\, dx^3$, \, $x^0 = c t$, 
	$F_{\mu\nu} = \partial_\mu \Afield_\nu - \partial_\nu \Afield_\mu$, 
	${\mathbf E}= (F_{01}, F_{02}, F_{03})$, and 
	${\mathbf B}= (F_{32}, F_{13}, F_{21})$. 
	
	Indices from the middle of the Greek alphabet ($\mu, \nu, \ldots$) 
	run from $0$ to $3$ and are raised and lowered
	with the spacetime metric $(g_{\mu\nu}) = {\rm diag}(1,-1,-1,-1)$. 
	In the four-potential, the temporal component $\Afield^0$ is the scalar potential 
	$\phi$, while the spatial components $\Afield^j$ ($j = 1, 2, 3$) form the 
	vector potential $\bAfield$. This is the standard Maxwell action, written here in the conventions used below; see, for example, Ref.~\cite{jackson1999classical}. 
	
	The interaction of a given atom with the EM field is described by 
	\begin{equation}\label{eq:SaI}
		\mathcal{S}_{I}^{a}({\mathbf r}_\alpha, \Afield ;  {\mathbf R}_\alpha ) =  
		\int dt  \,
		\left[\frac{q}{c} \, \dot{\mathbf r}_\alpha(t)\cdot \bAfield(t, 
		{\mathbf
			R}_\alpha + {\mathbf r}_\alpha(t)) -  q \, \phi(t, {\mathbf 
			R}_\alpha +
		{\mathbf r}_\alpha(t)) +  q\,  \phi(t, {\mathbf R}_\alpha) \right]\;,
	\end{equation} 
	where $q$ is the electron's charge. The first two terms are the standard
	nonrelativistic minimal coupling of the electron to the electromagnetic
	field, while the last term is the scalar-potential coupling of the fixed
	nucleus. The latter is needed because each atom is neutral: the electron has
	charge $-q$ at $\mathbf R_\alpha+\mathbf r_\alpha(t)$, and the nucleus has
	charge $+q$ at $\mathbf R_\alpha$. This form of the interaction is the
	starting point for the multipolar formulation used below
	\cite{cohen1989photons}.
	
	The electrons' dynamics is described nonrelativistically. It is worth
	noting that, up to this point, no dipole approximation has been 
	implemented.  
	
	
	\section{Coulomb gauge and instantaneous approximation}
	\label{sec:ham}
	
	\subsection{Coulomb gauge}
	
	The action in Eq.~(\ref{eq:defs}) corresponds to a Lagrangian
	$L$ (with ${\mathcal S} = \int dt\, L$) having the structure
	\begin{equation}\label{eq:Lagrangian_full}
		L = \frac{m}{2} \sum_\alpha \big( \dot{\mathbf r}_\alpha^2 - 
		\Omega^2   {\mathbf r}_\alpha^2  \big) +
		\int d^3x \Big[\frac{1}{2} \big( {\mathbf E}^2 - {\mathbf B}^2
		\big) - \frac{1}{c}  J_\mu(t, {\mathbf x}) \Afield^\mu(t,{\mathbf x}) \Big]\;,
	\end{equation}
	The first term in Eq.~(\ref{eq:Lagrangian_full}) is the mechanical part of the
	two atomic oscillators. The spatial integral contains the free Maxwell
	Lagrangian density and the coupling of the electromagnetic potential to the
	charge-current density. The current below is simply the sum of the electron
	current and the fixed nuclear charge contribution for each neutral atom.
	Explicitly, with the current four-vector $J^\mu$:
	\begin{align}
		J^0(t,{\mathbf x}) &= q  \sum_\alpha \Big[ 
		\delta^3\big({\mathbf x} - {\mathbf R}_\alpha - {\mathbf 
			r}_\alpha(t)\big) -
		\delta^3\big({\mathbf x} - {\mathbf R}_\alpha \big)\Big] \,c \equiv 
		\rho(t,{\mathbf x})  c \;, \nonumber\\
		{\mathbf J}(t,{\mathbf x}) & =  q  \sum_\alpha 
		\delta^3\big({\mathbf x} - {\mathbf R}_\alpha - {\mathbf 
		r}_\alpha(t)\big) 
		\dot{\mathbf r}_\alpha(t) \;.
	\end{align}
	Gauge invariance of the action is guaranteed by current conservation, 
	manifested as the continuity equation $\partial_\mu J^\mu = 0$. The total 
	charge $Q = \int d^3x\, \rho(t,{\mathbf x})$ vanishes, as does the 
	total charge of each atom.
	
	To proceed, we fix the gauge. Adopting the Coulomb gauge,
	$\nabla \cdot \bAfield = 0$, and noting that $\Afield^0$ is not
	dynamical, we can eliminate it from the Lagrangian by solving the equation it
	satisfies: $\nabla^2 \Afield^0 = - \rho$. This step separates the
	instantaneous Coulomb interaction, carried by the nondynamical scalar
	potential, from the transverse radiative degrees of freedom. Thus,
	\begin{align}\label{eq:L_coulomb}
		L &= \frac{m}{2} \sum_\alpha \big( \dot{\mathbf r}_\alpha^2 - 
		\Omega^2  {\mathbf r}_\alpha^2 \big)
		\nonumber\\
		& + \int d^3x \Big[\frac{1}{2} \big( \dot{\bAfield}_\perp^2 - {\mathbf 
		B}^2
		\big) + \frac{1}{c} {\mathbf J} \cdot \bAfield_\perp \Big]
		- U_{\rm Coul} \;,\nonumber\\
		U_{\rm Coul} &\equiv \frac{1}{2} \int d^3x \int d^3x'\, \rho(t,{\mathbf
			x})  \frac{1}{4 \pi |{\mathbf x} - {\mathbf x'}|}  \rho(t,{\mathbf 
			x'}) \;.
	\end{align}
	In Eq.~(\ref{eq:L_coulomb}) the term $U_{\rm Coul}$ contains all
	instantaneous electrostatic interactions among the electron and nuclear
	charges, including atom-independent self-energies. The remaining field
	variable is transverse. We have introduced the transverse part,
	$\bAfield_\perp$, of the vector potential:  
	\begin{equation}
		\bAfield_\perp(t, {\mathbf x}) =  \bAfield(t, {\mathbf x}) + 
		\nabla \int d^3x' \, \frac{1}{4 \pi |{\mathbf x} - {\mathbf x'}|}
		\nabla' \cdot \bAfield(t,{\mathbf x'}) \;,
	\end{equation}
	which satisfies $\nabla \cdot \bAfield_\perp = 0$. The same notation 
	shall 
	be used for any vector field to denote its solenoidal part. Analogously, we 
	shall use $\bAfield_\parallel$ for the longitudinal component,
	such that $\bAfield_\parallel = \bAfield - \bAfield_\perp$. The same convention will be used for the polarization field below: the subscript $\parallel$ denotes the longitudinal component and the subscript $\perp$ the transverse component.
	
	
	\subsection{The instantaneous approximation}
	\label{sec:inst}
	
	When the speeds involved in the electrons' motion are much smaller than $c$,
	one can neglect the coupling between the current and the vector potential.
	This is the instantaneous approximation: photons do not mediate retardation
	effects, and the only field remnant relevant for the atom-atom interaction
	is the Coulomb energy. The transverse part of the EM field becomes
	effectively decoupled from the atoms; thus, the effective Lagrangian used in
	this subsection is
	\begin{equation}
		L = \frac{m}{2} \sum_\alpha \big( \dot{\mathbf r}_\alpha^2 - 
		\Omega^2  {\mathbf r}_\alpha^2 \big)
		- U_{\rm Coul} \;.
	\end{equation}
	
	The Coulomb term may be written in terms of the 
	polarization\cite{cohen1989photons}. Since each atom contains two equal 
	and opposite charges, we have
	\begin{align}
		\delta^3\big({\mathbf x} - {\mathbf R}_\alpha - {\mathbf 
		r}_\alpha(t)\big) 
		-
		\delta^3\big({\mathbf x} - {\mathbf R}_\alpha\big) &=
		\int_0^1 du \, \frac{d}{du} \delta^3\big({\mathbf x} - {\mathbf 
		R}_\alpha -
		u {\mathbf r}_\alpha(t)\big) \nonumber\\
		&= -  \int_0^1 du \; {\mathbf r}_\alpha(t) \cdot \nabla  
		\delta^3\big({\mathbf x} - {\mathbf R}_\alpha -
		u {\mathbf r}_\alpha(t)\big) \nonumber\\
		&= -  \nabla  \cdot \int_0^1 du \; {\mathbf r}_\alpha(t) \,
		\delta^3\big({\mathbf x} - {\mathbf R}_\alpha -
		u {\mathbf r}_\alpha(t)\big) \;,
	\end{align}
	so that 
	\begin{equation}\label{eq:rho_p}
		\rho(t,{\mathbf x})= -\nabla \cdot {\mathbf P}(t,{\mathbf x})\;,
	\end{equation}
	with polarization ${\mathbf P} = \sum_\alpha {\mathbf P}_\alpha$ and 
	\begin{equation}
		{\mathbf P}_\alpha (t,{\mathbf x}) =  \int_0^1 du \; q 
		\, {\mathbf r}_\alpha(t) 
		\delta^3\big({\mathbf x} - {\mathbf R}_\alpha - u {\mathbf 
			r}_\alpha(t)\big) \;.
	\end{equation}
	
	Equation~(\ref{eq:rho_p}) allows us to express the Coulomb energy in 
	the alternative form
	\begin{equation}
		U_{\rm Coul} = \frac{1}{2}
		\int d^3x\, |{\mathbf P}_\parallel |^2 \;,
	\end{equation}
	where ${\mathbf P}_\parallel$ denotes the longitudinal, or irrotational,
	part of ${\mathbf P}$, defined by ${\mathbf P}_\parallel = {\mathbf P} -
	{\mathbf P}_\perp$ with $\nabla\cdot{\mathbf P}_\perp=0$. This notation
	is the same as that introduced above for the transverse and longitudinal
	parts of the vector potential.

	\subsection{Perturbative corrections to the vacuum energy}
	\label{ssec:pertub}
	
	We now quantize the instantaneous theory, since the following calculation
	uses ordinary perturbation theory around the oscillator ground state. The
	quantum version of this system yields London's expression for the 
	interaction energy between dipoles when the dipolar limit is taken, 
	but it also contains contributions from higher electric multipoles. The
	full quantum Hamiltonian $\hat{H}$ in the instantaneous approximation is
	\begin{equation}
		\hat{H}= \hat{H}_0 + \hat{H}'\;,
	\end{equation}
	with
	\begin{equation}
		\hat{H}_0 = \sum_{\alpha,i} \hbar \Omega \, 
		\big(\hat{a}_{\alpha,i}^\dagger 
		\hat{a}_{\alpha,i} + \tfrac{1}{2}\big) \;,
	\end{equation}	
	where $\hat{a}_{\alpha,i}$ and $\hat{a}_{\alpha,i}^\dagger$ are 
	annihilation and creation operators corresponding to each Cartesian 
	component of ${\mathbf r}_\alpha$:
	\begin{equation}\label{eq:r_operator}
		\hat{r}_\alpha^i = \sqrt{\frac{\hbar}{2 m \Omega}} 
		\big(\hat{a}_{\alpha,i}  +  \hat{a}_{\alpha,i}^\dagger \big) \;,\quad 
		[\hat{a}_{\alpha,i}, \hat{a}_{\beta,j}^\dagger] = 
		\delta_{\alpha\beta}\delta_{ij} \;,
	\end{equation}
	and 
	\begin{equation}
		\hat{H}' = \frac{1}{2} \int d^3x \int d^3x'\, \hat{\rho}(\mathbf{x}) \,
		\frac{1}{4\pi |\mathbf{x} - \mathbf{x}'|}\, \hat{\rho}(\mathbf{x}')\;.
	\end{equation}
	The perturbation $\hat{H}'$ contains self-energy contributions, which 
	do not affect the interaction between atoms. Setting 
	$\hat{H}' = \hat{H}_{\rm self} + \hat{H}_{\rm int}$ and 
	discarding $\hat{H}_{\rm self}$, we obtain
	\begin{equation}
		\hat{H}_{\rm int} = \int d^3x \int d^3x'\, \hat{\rho}_A(\mathbf{x}) \,
		\frac{1}{4\pi |\mathbf{x} - \mathbf{x}'|}\, \hat{\rho}_B(\mathbf{x}')\;,
	\end{equation}
	where
	\begin{equation}
		\hat{\rho}_\alpha(\mathbf{x}) = q \Big[\delta^3\big({\mathbf x} - 
		{\mathbf 
			R}_\alpha - \hat{\mathbf r}_\alpha\big) -\delta^3\big({\mathbf x} - 
		{\mathbf R}_\alpha\big) \Big] \;,\quad \alpha = A, B \;.
	\end{equation}
	
	We have split the Hamiltonian such that the unperturbed vacuum energy 
	$E_0^{(0)}$ is the sum of the zero-point energies of the two atoms:
	$E_0^{(0)} = 3 \hbar \Omega$. 
	The first-order correction $E_0^{(1)}$ is given by
	\begin{equation}
		E_0^{(1)}= \langle 0 | 	\hat{H}_{\rm int} | 0 \rangle \;,
	\end{equation}
	where $|0\rangle$ denotes the state annihilated by all 
	$\hat{a}_{\alpha,i}$.  
	This term vanishes, as can be seen by noting that it equals
	the classical Coulomb interaction energy with the charge densities replaced 
	by 
	their vacuum expectation values:
	\begin{equation}
		\langle 0 |\hat{\rho}_\alpha(\mathbf{x})| 0 \rangle = q \,
		\Big[ \Big(\frac{m \Omega}{\pi \hbar}\Big)^{3/2}\,	 e^{ - \frac{m 
				\Omega}{\hbar}({\mathbf x} - {\mathbf 
				R}_\alpha)^2 } -\delta^3\big({\mathbf x} - 
		{\mathbf R}_\alpha\big) \Big] \;.
	\end{equation}
	The spherical symmetry of the vacuum charge densities about each 
	nucleus implies $E_0^{(1)}=0$.
	
	The interaction Hamiltonian can be written equivalently in terms of the 
	polarization by integrating by parts:
	\begin{equation}
		\hat{H}_{\rm int} = \int d^3x \int d^3x' \, 
		\hat{P}_A^i(\mathbf{x}) \,F^{ij}({\mathbf x} -{\mathbf x}')\, 
		\hat{P}_B^j(\mathbf{x}')\;,
	\end{equation}
	with $F^{ij}(\mathbf{R}) = \frac{1}{4\pi}\left(\frac{3 R^i 
		R^j}{|\mathbf{R}|^5} - 
	\frac{\delta^{ij}}{|\mathbf{R}|^3}\right)$, where 
	$\langle 0 | \hat{\mathbf P}_\alpha | 0 \rangle = 0$. Note that we assume 
	no correlations between $\hat{\mathbf P}_A$ and $\hat{\mathbf P}_B$.
	
	Moving to the second-order term, $E_0^{(2)}$:  
	\begin{equation}
		E_0^{(2)} = -\sum_{n \neq 0} \frac{|\langle 0 | \hat{H}_{\rm int} | n 
			\rangle|^2}{E_n^{(0)} - E_0^{(0)}} \;.
	\end{equation}
	Using a multipole expansion of the polarization operator $\hat{\mathbf 
	P}$:  
	\begin{equation}
		\hat{\mathbf P}=\hat{\mathbf P}^{(2)} + \hat{\mathbf P}^{(4)}+
		\ldots\;,
	\end{equation} 
	where we follow the convention of using $2^{2k}$ to label the 
	corresponding terms (order $k=0$ means monopole, $k=1$ dipole, etc.),
	we obtain an infinite number of second-order contributions by 
	inserting the multipole expansion in $\hat{H}_{\rm int}$. The simplest one 
	arises from the dipolar approximation, where 
	$\hat{\mathbf P} \simeq \hat{\mathbf P}^{(2)}_\alpha$ 
	with $\hat{\mathbf P}^{(2)}_\alpha = \hat{\mathbf{d}}_\alpha \, 
	\delta^3({\mathbf x} - {\mathbf R}_\alpha)$ and 
	$\hat{\mathbf{d}}_\alpha = q\hat{\mathbf{r}}_\alpha$.  
	
	To second order, only 
	contributions from excited states of the form 
	\begin{equation}
		|n\rangle = |i\rangle_A |j\rangle_B \;, \quad i,  j  = 
		1, 2, 3 \quad \text{(dipole directions)}\;,
	\end{equation}
	survive, leading to
	\begin{equation}\label{eq:London}
		E_0^{(2)} = E_{\text{London}} = 
		-\frac{3}{4} \hbar \, \frac{\alpha^2\Omega}{R^6}\;,
	\end{equation}
	where the static polarizability is
	\begin{equation}\label{eq:polarizability}
		\alpha = \frac{q^2}{4\pi m\Omega^2}\;.
	\end{equation}
	This is London's renowned result, displaying the characteristic $R^{-6}$ 
	dependence. 
	
	In what follows we will often rewrite the energy denominators that appear 
	in 
	stationary perturbation theory using integral representations that 
	introduce an 
	\emph{auxiliary} time variable. By this device, including higher-order 
	terms 
	becomes more tractable. However, note that this is a bookkeeping device: 
	the 
	Hamiltonian remains time independent and no external driving is implied. 
	Retardation enters explicitly only through the electromagnetic-field correlators at finite separation in 
	Sec.~\ref{sec:ret}.
	
	\subsection{Introducing a time-dependent formulation for higher orders} 
	\label{sec:time_dep}
	
	Let us see how the previous result may be reformulated in a time-dependent 
	fashion. Since $\langle 0 | \hat{\mathbf P}_\alpha({\mathbf 
		x}) | 0\rangle = 0$ implies $\langle 0 | \hat{H}_{\rm int}| 0\rangle = 
		0$, 
	we can write $E_0^{(2)}$ as 
	\begin{equation}
		E^{(2)}_0 = - \langle 0 |\hat{H}_{\rm int} \, \frac{1}{:\!\hat{H}_0\!:} 
		\, 
		\hat{H}_{\rm int} |0 \rangle \;,
	\end{equation}
	with $:\!\hat{H}_0\!: \equiv \hat{H}_0 -\langle 0|\hat{H}_0|0\rangle$. 
	
	The key step is to use the integral representation
	\begin{equation}
		\frac{1}{:\!\hat{H}_0\!:} = \frac{1}{\hbar}\int^{\infty}_0 d\tau\, 
		e^{-\tau 
			:\hat{H}_0:/\hbar}\;,
	\end{equation}
	which allows us to write
	\begin{equation}
		E^{(2)}_0 = - \frac{1}{\hbar}\int^{\infty}_{0} d\tau\,  \langle 
		0 |\hat{H}_{\rm int} \, e^{-\tau :\hat{H}_0:/\hbar} \, \hat{H}_{\rm 
		int}
		|0 \rangle \;.
	\end{equation}
	This may be rewritten equivalently as
	\begin{align}
		E^{(2)}_0 & = - \lim_{T \to \infty} \left\{ \frac{1}{\hbar T} \,
		\int^{T}_{0} d\tau_1 \int_0^{\tau_1} d\tau_2   
		\; \langle 0 |\hat{H}_{\rm int} \, e^{-(\tau_1 - 
		\tau_2):\hat{H}_0:/\hbar} \, 
		\hat{H}_{\rm int} |0 \rangle  \right\} \nonumber\\
		& = - \lim_{T \to \infty} \left\{ \frac{1}{2 \hbar T} \,
		\int^{T}_{-T} d\tau_1 \int_{-T}^{\tau_1} d\tau_2   
		\; \langle 0 |e^{\tau_1 :\hat{H}_0:/\hbar} \hat{H}_{\rm int}  
		e^{-\tau_1 :\hat{H}_0:/\hbar} \,e^{\tau_2 :\hat{H}_0:/\hbar} 
		\hat{H}_{\rm int}   
		e^{- \tau_2:\hat{H}_0:/\hbar} |0 \rangle  \right\} \;.
	\end{align}
	Using the interaction representation in imaginary time, 
	$\hat{\mathcal O}(\tau) \equiv e^{\tau :\hat{H}_0:/\hbar}
	\hat{\mathcal O}  e^{-\tau :\hat{H}_0:/\hbar}$, and the usual time 
	ordering 
	$\mathrm{T}$: 
	\begin{equation}
		E^{(2)}_0 = - \lim_{T \to \infty} \left\{ \frac{1}{2\hbar T}\, 
		\int^T_{-T} d\tau_1  \int^T_{-T} d\tau_2\,
		\langle 0 |
		\mathrm{T} \big[ \hat{H}_{\rm int}(\tau_1) \hat{H}_{\rm 
		int}(\tau_2)\big] | 0 
		\rangle  \right\}\;.
	\end{equation}
	
	By an entirely analogous approach, the $n$th-order term (see the Appendix) 
	can be written as
	\begin{equation}
		E^{(n)}_0 = (-1)^{n-1} \lim_{T \to \infty} \left\{ \frac{1}{T\hbar^{n-1} 
		n !}\, 
		\int^T_{-T} d\tau_1  \ldots \int^T_{-T} d\tau_n\,
		\langle 0 |
		\mathrm{T} \big[ \hat{H}_{\rm int}(\tau_1) \ldots 
		\hat{H}_{\rm int}(\tau_n)\big] | 0 
		\rangle  \right\}_{\rm c}\;,
	\end{equation}
	where the subscript ``c'' indicates that only the 
	connected contribution should be considered.
	
	The sum of all orders can be rendered as
	\begin{align}
		E_0 &= \sum_{n=0}^\infty\, E^{(n)}_0 \nonumber\\
		& = - \lim_{T \to \infty}  \frac{\hbar}{T}
		\left[ \langle 0 |\sum_{n=0}^\infty \frac{(-1)^n}{n !}\, \mathrm{T}
		\Big( \frac{1}{\hbar}\int^T_{-T} d\tau\,  \hat{H}_{\rm int}(\tau) 
		\Big)^n | 0 \rangle  
		\right]_{\rm c} \nonumber\\
		& = - \lim_{T \to \infty} \frac{\hbar}{T} \left[ 
		\langle 0 |
		\mathrm{T} \Big( e^{- \frac{1}{\hbar}\int^T_{-T} d\tau\,  \hat{H}_{\rm 
		int}(\tau) } 
		\Big) | 0 \rangle  \right]_{\rm c} \nonumber\\
		& = - \lim_{T \to \infty}  \frac{\hbar}{T}
		\log {\mathcal Z} \;,
	\end{align}
	where 
	\begin{equation}
		{\mathcal Z} = \langle 0 |
		\mathrm{T} \Big( e^{- \frac{1}{\hbar}\int^T_{-T} d\tau\,  \hat{H}_{\rm 
		int}(\tau) } 
		\Big) | 0 \rangle\;,
	\end{equation}
	and the logarithm extracts the connected piece.
	
	Wick's theorem\cite{wick1950evaluation} allows one to evaluate ${\mathcal Z}$ to all orders,
	based on knowledge of the contractions:
	\begin{equation}\label{eq:contraction}
		\langle 0 | \mathrm{T} [ {\hat r}_\alpha^i(\tau) {\hat 
		r}_\beta^j(\tau') 
		]|0\rangle  =  \delta_{\alpha\beta} \, \delta^{ij}\,
		\frac{\hbar}{2 m \Omega} e^{- \Omega |\tau - \tau'|} \;.
	\end{equation}
	The resulting energy, including all terms, can be written as
	\begin{align}\label{eq:E_all_orders}
		E_0 &= \hbar  \int_0^\infty \frac{d\nu}{2 \pi} 
		\left\{ 
		2  \log\Big[ 1 - \Big(\frac{\alpha  \Omega^2}{(\nu^2 + \Omega^2) R^3 
		}\Big)^2 \Big]	\right. \nonumber\\
		&\left. +  \log\Big[ 1 - \Big(\frac{2 \alpha  \Omega^2}{(\nu^2 + 
			\Omega^2) 
			R^3 }\Big)^2 \Big]\right\} \;.
	\end{align}
	One can verify that London's result [Eq.~(\ref{eq:London})] emerges when
	$\alpha/R^3 \ll 1$, or equivalently, $q^2/R \ll m \Omega^2 R^2$.
	This condition states that the Coulomb interaction between an electron and 
	the other nucleus is much smaller than the energy binding it to its own.
	
	As explained in Ref.~\cite{Fosco:2023sbb}, for small values of $R$, 
	thresholds appear for the emergence of an imaginary part in the energy:
	\begin{equation}
		R_1 \equiv \Big( \frac{q^2}{2\pi m \Omega^2} \Big)^{1/3} \;,\quad
		R_2 \equiv \Big( \frac{q^2}{4\pi m \Omega^2} \Big)^{1/3} \;.
	\end{equation}
	This imaginary part signals a
	nonvanishing probability of vacuum decay: when the two atoms are 
	sufficiently 
	close to each other, the true vacuum should be closer to that of two 
	electrons in a molecule. 
	
	Here we have obtained this result in the context of the instantaneous 
	approximation, which is appropriate for describing this 
	short-distance phenomenon.
	
	
	\section{Including retardation}
	\label{sec:ret}
	
	\noindent\textbf{Units and notation.} Throughout this section we keep the 
	speed of light $c$ explicit in the discussion of scales, but we will set 
	$c=1$ in intermediate momentum-space steps so that the photon dispersion 
	reads $\omega_k = k$ and the Fourier variable $k\equiv|\boldsymbol{k}|$ may 
	be interpreted as an angular frequency. Restoring $c$ is then achieved by 
	replacing $k\to c\,k$ (and likewise $k'\to c\,k'$) and, equivalently, by 
	using the dimensionless combination $\rho\equiv \Omega R/c$ when presenting 
	results as functions of the separation $R$.

	At this point the content of the action changes in a controlled way: we keep
	the atomic oscillator actions and the free transverse electromagnetic field,
	but we no longer discard the transverse atom-field coupling. Assuming we 
	want
	to use the dipole approximation while keeping retardation effects, it is
	convenient to use the Power-Zienau-Woolley transformation
	\cite{power1959coulomb,woolley1971molecular}, whereby we add to $L$ a total
	derivative: $L \to L_F = L + \frac{d F}{dt}$ with
	$F(t) = - \int d^3x \, {\mathbf P}(t,{\mathbf x}) \cdot
	\bAfield_\perp(t,{\mathbf x})$.
	
	The resulting Lagrangian is
	\begin{equation}
		L_F = \frac{m}{2} \sum_\alpha\big( \dot{\mathbf r}_\alpha^2 - \Omega^2 
		{\mathbf r}_\alpha^2 \big) + \int d^3x \Big[\frac{1}{2} 
		\big( \dot{\bAfield}_\perp^2 - {\mathbf B}^2
		\big) + {\mathbf J}_M \cdot \bAfield_\perp -{\mathbf P} \cdot 
		\dot{\bAfield}_\perp \Big] - U_{\rm Coul} \;.
	\end{equation}
	where  ${\mathbf J}_M =  {\mathbf J} -  {\mathbf J}_P$, the latter 
	being the polarization current: ${\mathbf J}_P = \dot{\mathbf P}$. It is
	straightforward to show that
	\begin{equation}
		{\mathbf J}_M(t,{\mathbf x}) = \nabla \times {\mathbf
			M}(t,{\mathbf x}) \;,\quad
		{\mathbf M}(t,{\mathbf x})= q \sum_\alpha \int_0^1 du\, u \,
		{\mathbf r}_\alpha(t) \times \dot{\mathbf r}_\alpha(t) \,
		\delta^3\big({\mathbf x} - {\mathbf R}_\alpha - u {\mathbf
			r}_\alpha(t)\big) \;.
	\end{equation}
	
	After integrating by parts,
	\begin{equation}
		L_F =\frac{m}{2} \sum_\alpha\big( \dot{\mathbf r}_\alpha^2 - \Omega^2 
		{\mathbf r}_\alpha^2 \big) + \int d^3x \Big[\frac{1}{2} \big( 
		\dot{\bAfield}_\perp^2 - {\mathbf B}^2 \big) + {\mathbf M} \cdot {\mathbf B} 
		-{\mathbf P} \cdot \dot{\bAfield}_\perp \Big] - U_{\rm Coul} \;.
	\end{equation}
	Note that we could replace ${\mathbf M}$ and ${\mathbf P}$ by 
	${\mathbf M}_\perp$ and ${\mathbf P}_\perp$, respectively, 
	since they are multiplied by solenoidal fields. 
	
	The coupling between the EM field and the charges is contained in the 
	expression
	\begin{equation}
		L_I = \int d^3x \,({\mathbf M}_\perp \cdot {\mathbf B} 
		+{\mathbf P}_\perp \cdot  {\mathbf E}_\perp)\;.
	\end{equation}
	For a neutral system, the Coulomb energy may be written equivalently as
	\begin{equation}
		U_{\rm Coul} = \frac{1}{2} \int d^3x \,|{\mathbf P}_\parallel|^2\;.
	\end{equation}
	
	Neglecting contributions from ${\mathbf M}$ (setting ${\mathbf M}=0$), 
	the resulting approximate Lagrangian becomes
	\begin{equation}
		L =\frac{m}{2} \sum_\alpha\big( \dot{\mathbf r}_\alpha^2 - \Omega^2 
		{\mathbf r}_\alpha^2 \big) + \frac{1}{2} \int d^3x 
		\Big[\big(\dot{\bAfield}_\perp - {\mathbf P}_\perp \big)^2 - {\mathbf 
			B}^2 
		\Big] \;,
	\end{equation}
	where we completed a square 
	and used the property that 
	$\int d^3x \,|{\mathbf P}|^2$ contains only self-energy terms.
	
	Using ${\mathbf \Pi}_\perp$ ($= - {\mathbf E}_\perp$) to denote the 
	conjugate 
	momentum to $\bAfield_\perp$, we get for the classical Hamiltonian
	\begin{equation}
		H =\sum_\alpha\Big( \frac{{\mathbf p}_\alpha^2}{2 m} + 
		\frac{m \Omega^2 {\mathbf r}_\alpha^2}{2}   \Big) + \frac{1}{2} \int 
		d^3x 
		\Big[ {\mathbf \Pi}_\perp^2 +
		{\mathbf B}^2 + {\mathbf \Pi}_\perp \cdot {\mathbf P}_\perp \Big] \;.
	\end{equation}
	
	The Hamiltonian thus has the structure
	\begin{equation}
		\hat{H} = \hat{H}_{\rm atom} + \hat{H}_{\rm EM} + \hat{H}_I \;,   
	\end{equation}
	with the interaction $\hat{H}_I = - \int d^3x\, \hat{\mathbf P}_\perp \cdot 
	\hat{\mathbf E}_\perp$.
	
	The quantum Hamiltonian for this system can be written as
	\begin{equation}
		\hat{H} = \hat{H}_{\rm atom} + \hat{H}_{\rm EM} + 
		\hat{H}' \;,   
	\end{equation}
	where $\hat{H}_{\rm atom}$ describes the two harmonic oscillators, 
	$\hat{H}_{\rm EM}$ is the free electromagnetic field Hamiltonian, and 
	$\hat{H}'$ contains the atom-field coupling.
	
	Now consider the description of the system states, which are
	conveniently constructed as tensor products between states of each atom and 
	the electromagnetic field. The term $\hat{H}_0 = \hat{H}_{\rm atom} + 
	\hat{H}_{\rm EM}$ contains no 
	interaction and its eigenstates are known. In particular, the vacuum 
	consists of both atoms in the ground state together with the 
	electromagnetic 
	field vacuum (absence of photons).
	
	From now on we consider $\hat{H}_0$ normal-ordered, so the 
	energy of the ground state $| 0 \rangle = | 0 \rangle_{\text{EM}} | 0 
	\rangle_A 
	| 0 \rangle_B$ is zero. This convention simplifies the notation and does 
	not 
	affect the interaction potential between atoms, since that interaction only 
	involves energy differences.
	
	The first nontrivial order yielding a distance-dependent energy 
	is fourth order in charge. Discarding terms independent of the distance 
	between atoms, we have
	\begin{equation}\label{eq:E4_full}
		E^{(4)}_0 = - \sum_{n_1,n_2,n_3 \neq 0} \frac{\langle 0 |\hat{H}'| n_1 
			\rangle \langle n_1 |\hat{H}'| n_2 \rangle \langle n_2 |\hat{H}'| 
			n_3 \rangle 
			\langle n_3 |\hat{H}'| 0 \rangle }{\left( E^{(0)}_{n_1} - E^{(0)}_0 
			\right) 
			\left( E^{(0)}_{n_2} - E^{(0)}_0 \right) \left( E^{(0)}_{n_3} - 
			E^{(0)}_0 
			\right)}\;.
	\end{equation}
	For example, one set of intermediate states has the form
	\begin{align*}
		&|n_1\rangle = |k^i,\lambda\rangle_{\rm EM}|i\rangle_A|0\rangle_B\,, & 
		\Delta E^{(0)}_{n_1} = \hbar\left(\Omega + k\right)\;,\\
		&|n_2\rangle = |k^i,\lambda;k'^{j},\lambda'\rangle_{\rm EM}
		|0\rangle_A|0\rangle_B\,, & \Delta E^{(0)}_{n_2} = \hbar(k + k')\;,\\
		&|n_3\rangle = |k'^{j},\lambda'\rangle_{\rm 
		EM}|0\rangle_A|j\rangle_B\,, 
		& \Delta E^{(0)}_{n_3} = \hbar(\Omega + k')\;.
	\end{align*}
	The matrix elements take the form
	\begin{equation}
		\langle 0 |\hat{H}'|k^i,\lambda\rangle_{\rm EM}|i\rangle_A|0\rangle_B = 
		-\frac{q\hbar}{\sqrt{2m\Omega}}\frac{\mathrm{i}k}{\sqrt{(2\pi)^3 
				2k}}\varepsilon^i_{k\lambda}e^{-\mathrm{i}\mathbf{k}\cdot\mathbf{R}_A} 
				\;.
	\end{equation}
	Summing over all free indices, we use the property
	\begin{equation}
		\sum_{\lambda} \varepsilon^i_{k\lambda} \varepsilon^{j}_{k\lambda} = 
		\delta^{ij} - \frac{k^ik^j}{k^2} \;.
	\end{equation}
	Adding all 12 contributions, the interaction energy is obtained by 
	performing 
	the integral \cite{Power1957}
	\begin{align}\label{eq:E4_integral}
		E^{(4)}_0 =& - \frac{\hbar}{2} {\left(\frac{q^2}{2m\Omega}\right)}^2 \int 
		\frac{d\mathbf{k}}{(2\pi)^3} \int 
		\frac{d\mathbf{k}'}{(2\pi)^3}\, 
		\frac{(1+(\hat{k}\cdot\hat{k}')^2)\, kk'\, 
		e^{\mathrm{i}\mathbf{k}\cdot\mathbf{R}} 
			e^{\mathrm{i}\mathbf{k}'\cdot\mathbf{R}}}{(\Omega+k)(\Omega+k')}\nonumber\\
		&\times\left[ \frac{1}{\Omega(\Omega+k')} + \frac{1}{(k+k')(\Omega+k)} 
		+ 
		\frac{1}{(k+k')(\Omega+k')} + \frac{1}{(\Omega + k)(\Omega + k')} 
		\right]\;.
	\end{align}
	
	Let us now see how Eq.~(\ref{eq:E4_full}) can be simplified by following 
	the same approach as for the instantaneous approximation, now with 
	\begin{equation}\label{eq:Hprime_timedep}
		\hat{H}'(\tau) = -\hat{d}^i_A(\tau) \hat{E}^i (\boldsymbol{R}_A,\tau) 
		-\hat{d}^j_B(\tau) \hat{E}^j (\boldsymbol{R}_B,\tau) \;.
	\end{equation}
	Equation~(\ref{eq:Hprime_timedep}) is the imaginary-time interaction-picture
	version of the dipole Hamiltonian. The parameter $\tau$ is not an external
	time-dependent perturbation; it is the auxiliary Euclidean time generated by
	the unperturbed Hamiltonian and is introduced to rewrite the energy
	denominators as time-ordered correlators. Since $\hat{H}'$ is a sum of two terms, by distribution one would have to 
	calculate 16 expectation values and then sum them. Some vanish and others 
	give rise to self-energies; only 6 of those terms depend on the distance 
	between atoms. Each term is a product of four dipole moment and electric 
	field 
	operators. Wick's theorem allows us to separate the expectation value into 
	a 
	sum of products of contractions. These 
	contractions involve two operators acting in the same space. The 6 
	expectation values that depend on the interatomic distance have 2 
	possible contractions. In total, there are 12 terms contributing to the 
	interaction, all equal. An example is
	\begin{equation}
		\langle \hat{d}^{i_1}_A(\tau_1) \hat{d}^{i_2}_A(\tau_2) \rangle \langle 
		\hat{d}^{i_3}_B(\tau_3) \hat{d}^{i_4}_B(\tau_4) \rangle \langle 
		\hat{E}^{i_1} 
		(\boldsymbol{R}_A,\tau_1) \hat{E}^{i_3} (\boldsymbol{R}_B,\tau_3) 
		\rangle 
		\langle 
		\hat{E}^{i_2} (\boldsymbol{R}_A,\tau_2) \hat{E}^{i_4} 
		(\boldsymbol{R}_B,\tau_4) 
		\rangle\;.
	\end{equation}
	
	From now on, $\langle \hat{\mathcal{O}} \rangle = \langle 
	0|\mathrm{T}[\hat{\mathcal{O}}]|0 \rangle$ denotes the time-ordered vacuum 
	expectation value. Using the definition of the dipole moment operator 
	$\hat{\mathbf{d}}_\alpha = q\hat{\mathbf{r}}_\alpha$ 
	[cf.~Eq.~(\ref{eq:r_operator})] and its time evolution, we obtain
	\begin{equation}
		M^{ij}(\tau-\tau') = \langle \hat{d}^{i}_A(\tau) \hat{d}^{j}_A(\tau') 
		\rangle 
		= \delta^{ij} 
		\frac{q^2 \hbar}{2m\Omega}e^{-\Omega|\tau-\tau'|} = \hbar\delta^{ij} 
		\int^{\infty}_{-\infty} 
		\frac{d\nu}{2\pi}\, \tilde{M}(\nu)e^{i\nu (\tau-\tau')}\;,
	\end{equation}
	with
	\begin{equation}
		\tilde{M}(\nu) =  \frac{q^2}{m}\frac{1}{\nu^2 + \Omega^2}\;.
	\end{equation}
	
	On the other hand, according to Maxwell's equations, the electric field can 
	be 
	calculated from the four-potential:
	\begin{equation}
		\hat{E}^{i} (x) = \partial^{0}\hat{\Afield}^{i} (x) - \partial^{i}\hat{\Afield}^{0} 
		(x)\;,
	\end{equation}
	where $x$ is the four-vector $(t,\boldsymbol{x})$. This allows calculation 
	of 
	the contraction between two electric field operators from the photon 
	propagator. 
	Although the interaction was introduced in the Coulomb gauge, it is 
	convenient here to
	evaluate the gauge-invariant correlator 
	$I^{ij}=\langle\hat{E}^i\hat{E}^j\rangle$ in a covariant (Feynman) gauge; 
	the final result for $I^{ij}$ is gauge independent.
	In the imaginary-time scheme, the propagator takes the form
	\begin{equation}
		\langle \hat{\Afield}^{\mu} (x) \hat{\Afield}^{\nu} (x') \rangle = \hbar\delta^{\mu\nu} 
		\int 
		\frac{d^4k}{(2\pi)^4}\, \frac{e^{ik\cdot(x-x')}}{k^2}\;.
	\end{equation}
	
	In this imaginary-time scheme (Wick rotation $t\to -i\tau$), the Minkowski 
	metric $g^{\mu\nu}$ is replaced by the Euclidean metric $\delta^{\mu\nu}$. 
	We write the Euclidean four-momentum as $k=(\nu,\boldsymbol{k})$, with 
	$k^2=\nu^2+\boldsymbol{k}^2$. Fourier-transforming the derivatives in 
	$\hat{E}^i=\partial^0\hat{\Afield}^i-\partial^i\hat{\Afield}^0$ then yields, after some 
	algebra
	\begin{equation}
		I^{ij} (x-x') = \langle \hat{E}^{i} (x) \hat{E}^{j} (x') \rangle = 
		\hbar\delta^4(x-x') + \hbar\int 
		\frac{d^4k}{(2\pi)^4}\, \frac{e^{-ik\cdot(x-x')}}{k^2} \left( 
		-\delta^{ij} 
		\boldsymbol{k}^2 + k^i k^j \right)\;.
	\end{equation}
	
	We can discard the delta since the positions of the atoms never coincide. 
	This 
	integral can also be obtained by summing different derivatives with respect 
	to 
	$x$ of the photon propagator:
	\begin{equation}
		I^{ij} (x-x') = \hbar\left( \delta^{ij} \nabla^2 - \partial^i \partial^j 
		\right)\int \frac{d^4k}{(2\pi)^4}\, \frac{e^{-ik\cdot(x-x')}}{k^2}\;.
	\end{equation}
	
	It is possible to perform the spatial integral to obtain the 
	propagator in terms of a single integral:
	\begin{equation}
		I^{ij} (x-x') = \hbar\int^{\infty}_{-\infty} \frac{d\nu}{2\pi}\, 
		\tilde{I}^{ij}(|\boldsymbol{x}-\boldsymbol{x}'|,\nu)e^{i\nu 
		(\tau-\tau')}\;,
	\end{equation}
	with
	\begin{equation}
		\tilde{I}^{ij}(|\boldsymbol{x}-\boldsymbol{x}'|,\nu) = \left( 
		\delta^{ij} 
		\nabla^2 - \partial^i \partial^j \right) 
		\frac{e^{-|\nu||\boldsymbol{x}-\boldsymbol{x}'|}}{4\pi|\boldsymbol{x}-
			\boldsymbol{x}'|}\;.
	\end{equation}
	
	Putting all this together, the fourth-order energy correction becomes
	\begin{equation}
		E_I = - \frac{1}{\mathcal{N}}\frac{1}{\hbar^3}\frac{12}{4!} \left( \prod^4_{s=1} 
		\int^{\infty}_{-\infty} d\tau_s\right) M^{i_1 i_2} (\tau_1-\tau_2) 
		M^{i_3 i_4} 
		(\tau_3-\tau_4) 
		I^{i_1 i_3}(\boldsymbol{R},\tau_1 - \tau_3) I^{i_2 
		i_4}(\boldsymbol{R},\tau_2 - 
		\tau_4)\;.
	\end{equation}
	
	After some algebraic manipulations, we obtain
	\begin{equation}
		E_I = -\frac{1}{2}\frac{1}{\hbar^3} \left( \prod^3_{s=1} \int^{\infty}_{-\infty} 
		d\tau_s\right) M (\tau_1) M (\tau_3) I^{ij}(\boldsymbol{R},\tau_1 
		+\tau_2) 
		I^{ij}(\boldsymbol{R},\tau_2 + \tau_3)\;.
	\end{equation}
	
	Using that $\tilde{M}(\tau)=\tilde{M}(-\tau)$ and 
	$I^{ij}(\boldsymbol{R},-\tau)=I^{ij}(\boldsymbol{R},\tau)$, together with 
	the convolution theorem, we arrive at
	\begin{equation}\label{eq:E_I_frequency}
		E_I = -\frac{\hbar}{2} \int^{\infty}_{-\infty} \frac{d\nu}{2 \pi} 
		\left( 
		\tilde{M}(\nu)\tilde{I}^{ij}(\boldsymbol{R},\nu) \right)^2\;.
	\end{equation}
	
\begin{figure}[htb] 
		\centering
		\includegraphics[width=0.6\linewidth]{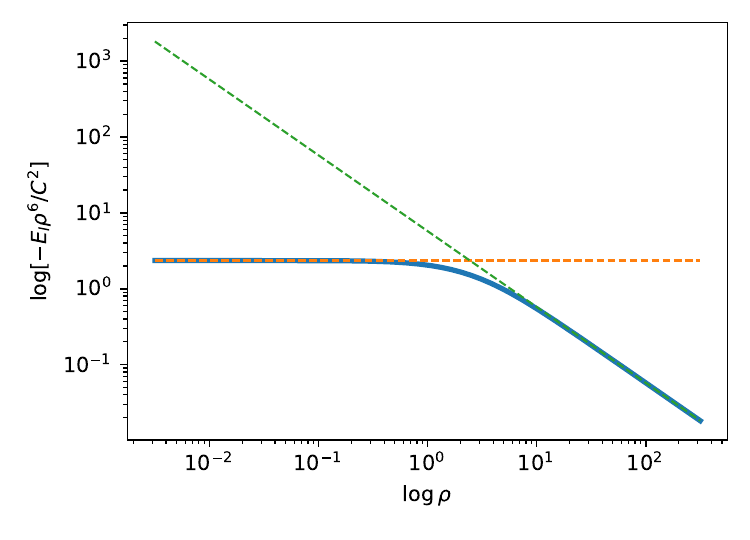}
		\caption{Numerical evaluation of Eq.~(\ref{eq:E_I_result}) shown on a 
		log-log scale. The thick curve plots the dimensionless combination 
		$-E_I(\rho)\, \rho^6/C^2$ versus $\rho\equiv\Omega R/c$, where $C\equiv 
		q^2\Omega/(4\pi m)$. Thin dashed lines show the expected behavior if 
		$E_I$ followed pure power laws: constant for $E_I \propto \rho^{-6}$ 
		(London regime, short distances) and $\propto \rho^{-1}$ for 
		$E_I \propto \rho^{-7}$ (Casimir-Polder regime, large distances).}
		\label{fig:E4}
	\end{figure}

\begin{figure}[htb] 
		\centering
		\includegraphics[width=0.6\linewidth]{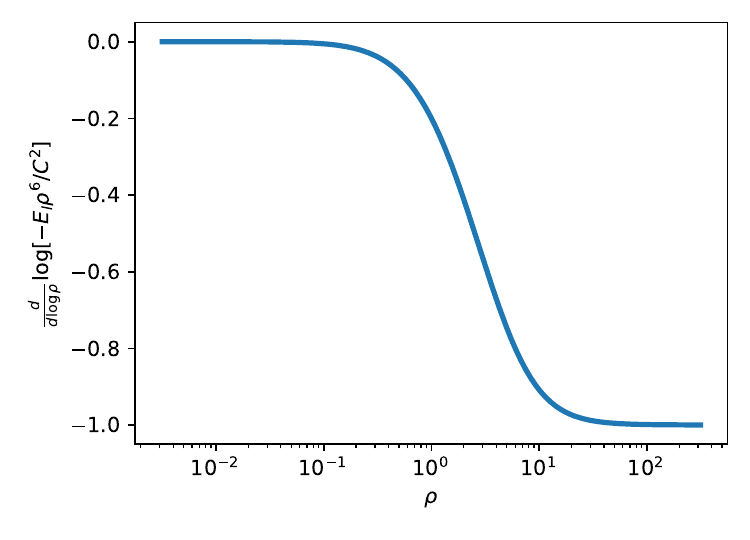}
		\caption{Logarithmic slope extracted from Fig.~\ref{fig:E4}: 
		$\mathrm{d}\log[-E_I(\rho)\,\rho^6]/\mathrm{d}\log \rho$. The curve interpolates 
		between the London exponent $-6$ for $\rho\ll 1$ and the Casimir-Polder 
		exponent $-7$ for $\rho\gg 1$.} 
		\label{fig:dE4}
	\end{figure}

	Summing over $i,j$ we obtain
	\begin{equation}\label{eq:E_I_result}
		E_I = -\frac{\hbar \Omega}{\pi} \left(\frac{q^2 \Omega}{4 \pi 
		m}\right)^2 
		\frac{1}{\rho^3} \int^{\infty}_{0} d\nu\, \frac{\nu^4+2\nu^3 + 5\nu^2 + 
		6\nu + 
			3}{(\nu^2+\rho^2)^2}\, e^{-2\nu}\;,
	\end{equation}
	where $\rho \equiv \Omega R/c$ (we used $c=1$ in intermediate steps). This symbol is used only for the dimensionless separation, while $\mathbf r_\alpha$ always denotes the internal electron coordinate of atom $\alpha$. This 
	result is equivalent to that obtained 
	in Ref.~\cite{Fosco:2023sbb}. Figure~\ref{fig:E4} shows 
	$-E_I \cdot \rho^6$, divided by $C^2$ with $C = q^2\Omega/4\pi m$, as a 
	function 
	of $\rho$ on a log-log scale. 
	Figure~\ref{fig:dE4} shows the numerical derivative of this curve, 
	revealing the exponent throughout the entire range of distances.
	
	Both asymptotic behaviors can be determined analytically. For 
	distances such that $\rho \ll 1$, the exponential decay ensures that $\nu$ 
	remains of order unity, so $\rho\nu \ll 1$, yielding
	\begin{equation}
		E_I \approx -3\hbar \left(\frac{q^2}{4 \pi m}\right)^2 \frac{1}{R^6} 
		\int^{\infty}_{0} \frac{d\nu}{2\pi}\, \frac{1}{(\nu^2+\Omega^2)^2} = 
		-\frac{3}{4} 
		\left( \frac{q^2}{4\pi m \Omega} \right)^2 \frac{\hbar}{\Omega R^6}\;.
	\end{equation}
	This result is identical to London's equation~(\ref{eq:London}). 
	
	The integral in Eq.~(\ref{eq:E_I_result}) can be evaluated exactly:
	\begin{equation}
		E_I = -\frac{\hbar\Omega}{2 \pi } \left(\frac{q^2 \Omega}{4 \pi 
		m}\right)^2 
		\frac{1}{\rho^6} \left[ \rho(6-\rho^2)+(3-7\rho^2+\rho^4)f(2\rho)+2\rho(3-3\rho^2+\rho^4)g(2\rho) 
		\right]\;,
	\end{equation}
	where
	\begin{align}
		f (x)&= \mathrm{Ci}(x)\sin{x}-\big(\mathrm{Si}(x)-\tfrac{\pi}{2}\big)
		\cos{x}\;,\\
		g (x)&= 
		-\big[\mathrm{Ci}(x)\cos{x}+\big(\mathrm{Si}(x)-\tfrac{\pi}{2}\big)
		\sin{x}\big]\;,
	\end{align}
	and $\mathrm{Ci}(x)$ and $\mathrm{Si}(x)$ are the cosine and sine 
	integrals, 
	respectively. For $\rho \gg 1$, the energy approaches
	\begin{equation}\label{eq:Casimir_Polder}
		E_I \approx - \frac{23\hbar c}{4\pi}\,\alpha^2\,\frac{1}{R^7}\;.
	\end{equation}
	
	Using the polarizability $\alpha = q^2/(4\pi m \Omega^2)$ 
	[Eq.~(\ref{eq:polarizability})], we can identify the factor that 
	corrects the London result:
	\begin{equation}
		E_I \approx - \frac{23\hbar c}{4\pi}\,\frac{\alpha^2}{R^7}
		= E_{\text{London}}\times \frac{23}{3\pi}\,\frac{c}{\Omega R}\;.
	\end{equation}

	It is useful to spell out how the instantaneous London limit is recovered
	within this correlator formulation. The required approximation is that the
	electric field propagator reduces to a delta function in time. This 
	approximation can be consistently applied at any stage of the calculation, 
	always yielding the same result. This can be understood as follows: 
	the harmonic oscillator propagator depends only on its energy, while the 
	electric field propagator depends only on the interatomic distance. 
	Effectively, the interaction becomes instantaneous 
	when the particles are sufficiently close, in which case we are essentially 
	calculating the Fourier transform of a delta function. From an operator 
	perspective, this requires the replacement
	\begin{equation}
		I^{ij} (x-x') \to 
		\tilde{I}^{ij}(\boldsymbol{R},0)\,\delta(\tau-\tau')\;,
	\end{equation}
	where
	\begin{equation}
		\tilde{I}^{ij}(\boldsymbol{R},0) = \left( \delta^{ij} \nabla^2 - 
		\partial^i 
		\partial^j \right) \frac{1}{4\pi R} = \frac{1}{4\pi }\left( 
		\frac{\delta^{ij}}{R^3} - \frac{3R^i R^j}{R^5}\right) - 
		\delta^{ij}\delta^{(3)}(\boldsymbol{R})\;,
	\end{equation}
	and thus 
	\begin{equation}
		\tilde{I}^{ij}(\boldsymbol{R},0)\tilde{I}^{ji}(\boldsymbol{R},0) = 
		\frac{6}{(4\pi)^2 R^6}\;.
	\end{equation}
	
	Although the results obtained in this work are 
	identical to those of Casimir and Polder, the approaches considered differ. 
	Casimir and Polder considered the retarded potential of 
	an atom that emits radiation and showed that, when retardation is 
	negligible, the London result is recovered. Our starting point was a model 
	with electrostatic and instantaneous interaction. Although we have 
	introduced 	an imaginary-time dependence, the perturbations employed were 
	time-independent. This highlights the static and instantaneous character of 
	the interaction, which is not obvious at first sight.

	
	\section{Conclusions}\label{sec:conc}
	
	We have presented a derivation of the van der Waals interaction energy 
	between two atoms, using first standard stationary perturbation theory as a 
	starting point, and then showing how a time-dependent formalism provides a 
	much simpler tool for evaluating higher orders.
	
	In the instantaneous approximation, we showed that when the 
	atoms are too close, an instability phenomenon, already predicted 
	in Ref.~\cite{Fosco:2023sbb} in the context of a path integral 
	calculation, is captured by our treatment. It should be clear, however, 
	that this instability signals the breakdown of our description; a 
	different framework (for instance, molecular orbitals) is needed in that 
	regime.
	
	For the long-distance case, where retardation effects are essential, 
	we demonstrated how the time-dependent treatment yields a much 
	simpler way to evaluate the interaction energy.   
	In this way, the same framework makes transparent how the London and
	Casimir-Polder regimes arise from different limits of the same
	correlator expression. The resulting formulation may be useful as advanced
	expository material for graduate students or specialized readers interested
	in the connection between stationary perturbation theory, correlation
	functions, and retarded dispersion forces.

\section*{Acknowledgments}
This work was supported by ANPCyT, CONICET, and UNCuyo.

	\appendix
	
	\section{Vacuum energy in Rayleigh-Schr\"odinger perturbation theory}
	\label{app:RSvacuum}
	
	We summarize here some relevant results on stationary perturbation theory, 
	following the resolvent approach of Refs.~\cite{kato1995perturbation,sakurai2017modern}, 
	as applied to the present context. 
	One starts by introducing a parameter $\lambda$ and considering the family 
	of 
	self-adjoint Hamiltonians
	\begin{equation}
		\hat H(\lambda) = \hat H_0 + \lambda  \hat H_{\rm int} \;,
	\end{equation}
	such that $\hat H = \hat H(1)$, where $\hat H_0$ has a nondegenerate 
	isolated vacuum with zero energy.
	We denote by $E(\lambda)$ the eigenvalue of $\hat H(\lambda)$ that is 
	analytic 
	near $\lambda=0$ and satisfies $E(0)=0$. We also introduce the projectors
	\begin{equation}
		\hat P \equiv |0\rangle\langle 0| \;, \quad \hat Q \equiv \hat I - \hat 
		P 
		\;,
	\end{equation}
	where $\hat I$ denotes the identity operator.
	
	Under these assumptions, there exists a positively oriented contour 
	$\Gamma$ enclosing $E(\lambda)$ (and no other point of the spectrum), and 
	one can write
	\begin{equation}
		E(\lambda)=\frac{1}{2\pi i}\oint_{\Gamma} z\,\mathrm{Tr}\!\left[ 
		\big( z-\hat H(\lambda) \big)^{-1}\right] dz \;.
		\label{eq:Efromresolventtrace}
	\end{equation}
	For $z\in\Gamma$ and $|\lambda|$ small, we have the Neumann series
	\begin{equation}
		(z-\hat H(\lambda))^{-1}
		=(z- \hat H_0)^{-1}\sum_{n=0}^{\infty} \left[\lambda 
		\hat H_{\rm int} (z- \hat H_0)^{-1}\right]^n \;.
		\label{eq:Neumann}
	\end{equation}
	We can then perform a Laurent expansion around $z=0$:
	\begin{equation}
		(z-\hat H_0)^{-1}=\frac{1}{z} \hat P - \sum_{k\ge 0} z^{k}\,\hat 
		S^{k+1} 
		\;,
		\label{eq:R0Laurent}
	\end{equation}
	where $\hat S^0 \equiv - \hat P$ and $\hat S^k \equiv \hat Q \, 
	\hat H_0^{-k}\, \hat Q$ for $k\ge 1$. The proof follows from noting that 
	on $\mathrm{Ran}(\hat P)$ one has $(z- \hat H_0)^{-1}=z^{-1} 
	\hat P$, while on $\mathrm{Ran}(\hat Q)$:
	\begin{equation}
		(z-\hat H_0)^{-1}=- \hat H_0^{-1}(1-z \hat H_0^{-1})^{-1}=-\sum_{k\ge 
		0} z^k\, 
		\hat Q \hat H_0^{-(k+1)} \hat Q \;,
	\end{equation}
	where $\hat H_0^{-1}$ is understood as the bounded inverse of $\hat H_0$ on 
	$\mathrm{Ran}(\hat Q)$ (which exists because $0$ is isolated).

	When inserted into Eq.~(\ref{eq:Efromresolventtrace}), this yields a 
	power series $E(\lambda)=\sum_{n\ge 1}\lambda^n E^{(n)}$, where $E^{(n)}$ 
	is 
	determined by extracting from
	Eq.~(\ref{eq:Efromresolventtrace}) the coefficient of $\lambda^n$ and then 
	taking the residue at $z=0$. Using Eq.~(\ref{eq:R0Laurent}),
	each factor $(z-\hat H_0)^{-1}$ contributes either $(1/z)\hat P$ or a 
	regular 
	term $-z^{k} \hat S^{k+1}$. The coefficient of $z^{-1}$
	arises from choosing exactly \emph{one} $(1/z) \hat P$ overall, while the 
	remaining 
	$(z-\hat H_0)^{-1}$ factors contribute
	regular powers whose total power of $z$ cancels the extra $z$ in 
	Eq.~(\ref{eq:Efromresolventtrace}).
	
	A convenient way to organize the resulting combinatorics is to record the 
	regular contributions via the operators $ \hat S^k$.
	The outcome is a closed formula for 
	$n\ge 1$, giving the $n$th-order vacuum-energy coefficient 
	as
	\begin{equation}
		E^{(n)} = \sum_{\substack{k_1,\dots,k_{n-1}\ge 0\\ 
		k_1+\cdots+k_{n-1}=n-1}} 
		\langle 0|\, \hat H_{\rm int}  \hat S^{k_1} \hat H_{\rm int} \hat 
		S^{k_2}\cdots 
		\hat H_{\rm int} \hat S^{k_{n-1}} \hat H_{\rm int}  \,|0\rangle\;,
		\label{eq:KatoEnergyCoeff}
	\end{equation} 
	with $\hat S^0 = - \hat P$ and $\hat S^{k\ge 1}= \hat Q \hat H_0^{-k} \hat 
	Q$.
	
	Equation~(\ref{eq:KatoEnergyCoeff}) produces, 
	through the presence of $\hat S^0 = - \hat P$, all the 
	``subtraction'' terms that appear in the usual stationary  
	Rayleigh-Schr\"odinger expressions (except for cyclic ones).
	
	Using $\hat S^1 = \hat Q \hat H_0^{-1} \hat Q$ and $\hat S^0 = - 
	\hat P$, subtraction terms first arise at fourth order.
	With $k_1+k_2+k_3=3$, one finds
	\begin{equation}
		E^{(4)}
		=\langle 0|\hat H_{\rm int} \hat S^{1}\hat H_{\rm int}  
		\hat S^{1} \hat H_{\rm int} \hat S^{1}\hat H_{\rm int} |0\rangle
		-\langle 0|\hat H_{\rm int}  \hat S^{1} \hat H_{\rm int} |0\rangle \, 
		\langle 0| \hat H_{\rm int} \hat S^{2} \hat H_{\rm int} |0\rangle\;,
		\label{eq:E4_appendix}
	\end{equation}
	where the second term arises from the contribution with the vacuum 
	projector 
	in the \emph{middle},
	\eg, $\langle 0|\hat H_{\rm int} \hat S^{1} \hat H_{\rm int} 
	\hat S^{0} \hat H_{\rm int}  \hat S^{2} \hat H_{\rm int} 
	|0\rangle=-\langle 0|\hat H_{\rm int} \hat S^{1} \hat H_{\rm int} 
	|0\rangle\langle 0| \hat H_{\rm int} \hat S^{2} \hat H_{\rm int} 
	|0\rangle$. This is precisely the familiar Rayleigh-Schr\"odinger 
	subtraction 
	structure at fourth order.
	
	The role of the subtraction terms is, in the general case, to make
	the correction at each order \emph{connected}, \ie, without any 
	sub-cycle (from vacuum to vacuum) like the one being subtracted
	in the second term of Eq.~(\ref{eq:E4_appendix}).

\end{document}